\documentclass[aip,apl,twocolumn,amsmath,amssymb,bibnotes]{revtex4}

\usepackage{color,epsfig,rotating,graphicx,subfigure}

\usepackage{amsmath,amssymb,amscd,ulem}

\usepackage[latin1]{inputenc}
\usepackage[english]{babel}

\usepackage{dsfont}
\renewcommand{\Re}{{\rm Re}}
\renewcommand{\Im}{{\rm Im}}

\newcommand{\ri}{{\rm i}}
\newcommand{\re}{{\rm e}}
\newcommand{\rd}{{\rm d}}

\newcommand{\Tr}{{\rm Tr}}

\begin{document}
\title{Radiative thermal diode driven by non-reciprocal surface waves}

\date{\today}

\author{Annika Ott$^1$, Riccardo Messina$^2$, Philippe Ben-Abdallah$^2$, Svend-Age Biehs$^{1}$}
\affiliation{$^1$ Institut f\"{u}r Physik, Carl von Ossietzky Universit\"{a}t, D-26111 Oldenburg, Germany}
\affiliation{$^2$ Laboratoire Charles Fabry,UMR 8501, Institut d'Optique, CNRS, Universit\'{e} Paris-Sud 11,
2, Avenue Augustin Fresnel, 91127 Palaiseau Cedex, France}

\begin{abstract}  
We demonstrate the possibility to rectify the nanoscale radiative heat flux between two nanoparticles by coupling them with the nonreciprocal surface modes of a magneto-optical substrate in a Voigt configuration. When the non-reciprocal medium supports a surface wave in the spectral window where heat exchanges take place the rectification coefficient can reach large values opening so the way to the design of true thermal diodes.
\end{abstract}


\maketitle

A thermal diode is a two-terminal solid element that conducts heat flux primarily in one direction. In other words it 
is a device which displays a highly asymmetric thermal conductance. This asymmetry generally results from a non-linear 
dependence of certain physical properties with respect to the temperature. To date numerous solid-state thermal diodes 
have been developed to rectify the heat carried by phonons by exploiting various physical mechanisms including nonlinear 
atomic vibrations~\cite{LiEtAl2004}, nonlinearity of the electron gas dispersion relation in metals~\cite{SegalEtAl2008}, 
direction dependent Kapitza resistances~\cite{CaoEtAl2012}, or dependence of the superconducting density of states and 
phase dependence of heat currents flowing through Josephson junctions~\cite{PerezEtAl2013}. More recently radiative 
thermal diodes have been proposed~\cite{PBASAB2013,ItoEtAl,FiorinoEtAl2018} to rectify the heat flux carried by thermal 
photons~\cite{FanRectification2010,BasuEtAl2011,WangEtAl2013,OrdonezEtAl2017} both in near-field (close separation) and 
in far-field (large separation) regimes. These devices are basically made with materials which have dielectric properties 
which strongly depend on the temperature. This situation occur for instance with phase change materials~\cite{QazilbashEtAl2007} 
such as metal-insulator transition materials whose the optical properties undergo a real bifurcation around their critical 
temperature. Although these devices display relatively high rectification coefficients in near-field the scarcity of these 
phase change materials could limit the development and the operating range of this technology. 

Here we explore a new route to rectify the radiative flux by using magneto-optical materials which have already shown interesting effects on near-field thermal
radiation between magneto-optical nanoparticles~\cite{Zhu2016,Latella2017,Cuevas,Ben-Abdallah2016,OttEtAl2018} as reviewed 
recently~\cite{OttEtAl2019}. Under the action of an external magnetic field these materials becomes non-reciprocal. 
In this Letter we show that the heat 
transfers between two nanoparticles which are placed close to these anisotropic and non-reciprocal media strongly depend on 
the sign of temperature gradient. We demonstrate that the thermal rectification is closely related to a 
strong asymmetry in the propagation of surface waves supported by these media. 

\begin{figure}
	\centering
	\includegraphics[width=0.4\textwidth]{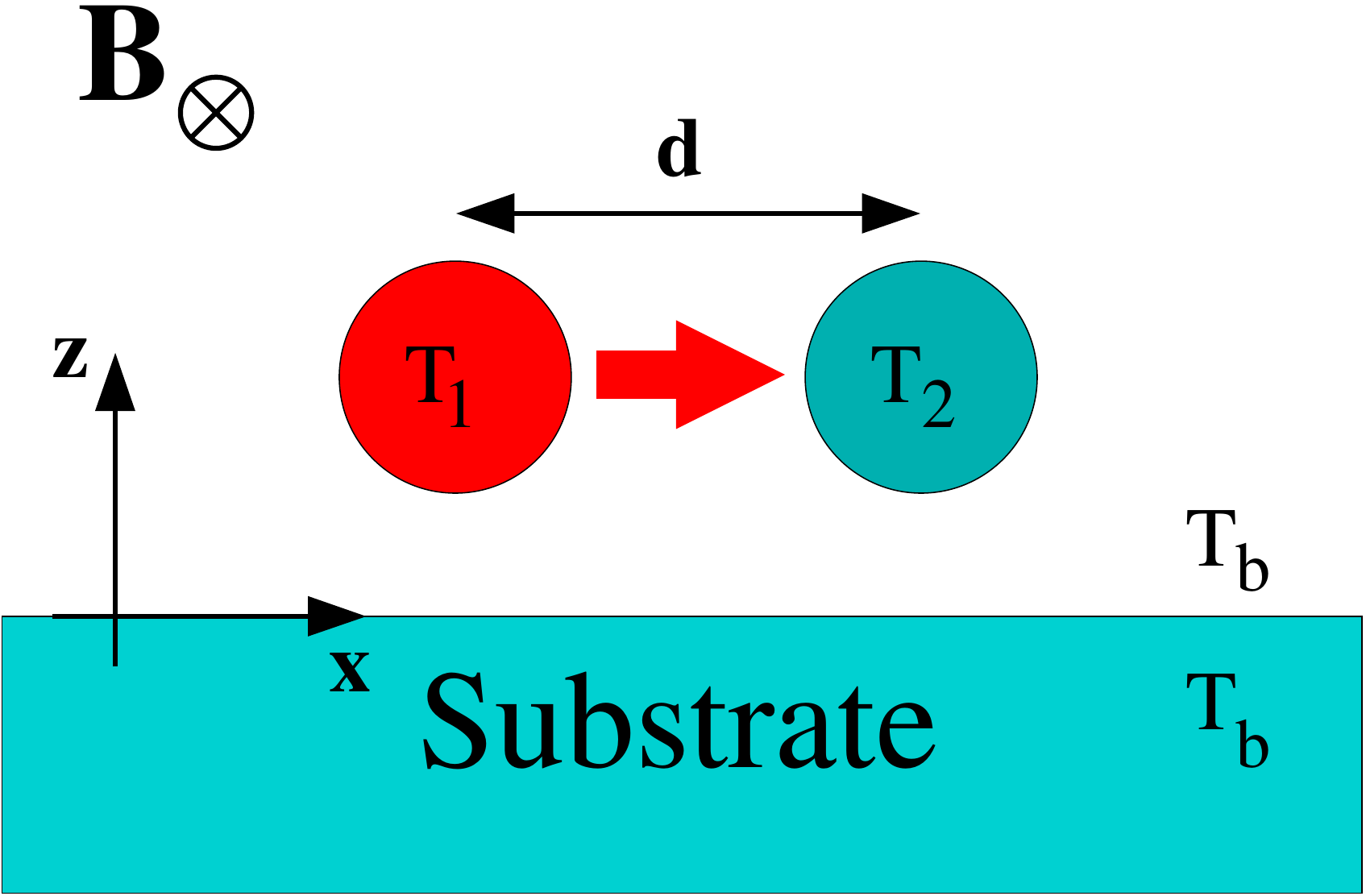}
	\caption{Sketch of the geometry. Two magneto-optical nanoparticles (InSb) are placed close to a magneto-optical substrate (InSb).
                 The magnetic field is aligned along the y axis (Vogt configuration). The two particles have temperatures $T_1$ and $T_2$ and the 
                 background has temperature $T_b$. In the ``forward'' case with $T_1 = T_p > T_2 = T_b$ we are interested in the heat flux
                 $\phi_F$ from particle one to particle two. Similarly, we can define $\phi_B$ as the heat flux from particle two to particle one
                 in the ``backward'' case with $T_2 = T_p > T_1 = T_b$.}
	\label{Fig:Geometry}
\end{figure}

We consider a setup as sketched in Fig.~\ref{Fig:Geometry} where two spherical InSb nanoparticles (labeled with 1 and 2) are at a distance $z$ from a planar InSb substrate. The two particles have temperatures $T_1$ and $T_2$, respectively, and the interparticle distance is $d$. The background consisting of the photon gas above the substrate and the substrate itself are assumed to have the temperature $T_b$. In this configuration the total power received by particle $i = 1,2$ can be written as ($j = 1,2; j \neq i$)
\begin{equation}
\begin{split}
  P_i &= 3 \int_0^\infty \!\! \frac{\rd \omega}{2 \pi}\, \bigl( \bigl[\Theta(T_j) - \Theta(T_i)\bigr] \mathcal{T}_{ji} \\
      &  \quad\quad\quad\quad + \bigl[\Theta(T_b) - \Theta(T_i)\bigr] \mathcal{T}_{bi} \bigr) 
\end{split}
\end{equation}
where $\Theta(T) = \hbar \omega / (\exp(\hbar \omega/ k_B T) - 1)$ is the thermal part of the mean energy of a harmonic oscillator; $\hbar$
is the reduced Planck contant and $k_B$ is the Boltzmann constant. $\mathcal{T}_{ji}$ and $ \mathcal{T}_{bi}$ are the transmission coefficients for the power exchanged between the particles and particle $i$ and the background. 

Now, we consider the following two cases: (i) the ``forward'' case, where we heat particle 1 to $T_p$ with respect to its environment so that $T_1 = T_p > T_2 = T_b$. Then the power received by particle $2$ is fully determined by $\mathcal{T}_{12}$. (ii) the ``backward'' case, where we heat particle 2 to $T_p$ with respect to its environment so that $T_2 = T_p > T_1 = T_b$. Then the power received by particle $1$ is fully determined by $\mathcal{T}_{21}$. Therefore, any assymetry in the heat flux, i.e.\ a rectification, can only exist if $\mathcal{T}_{12} \neq \mathcal{T}_{21}$. The expressions for the transmission coefficients can be derived from fluctuational electrodynamics theory and are in single-scattering approximation given by ($i,j = 1,2; i \neq j$)~\cite{EkerothEtAl2017}
\begin{align}
   \mathcal{T}_{ji} &= \frac{4}{3} k_0^4 \Tr\bigl[ \uuline{\chi} \mathds{G}_{ij} \uuline{\chi} \mathds{G}_{ij}^\dagger \bigr],
\end{align}
where we have introduced the wavenumber in vacuum $k_0 = \omega/c$. When neglecting the radiative correction we can write the response function of the nanoparticles $\uuline{\chi}$ in terms of the polarizability tensor $\uuline{\alpha}$ as~\cite{EkerothEtAl2017}
\begin{align}
    \uuline{\chi} = \frac{1}{2 \ri} \bigl( \uuline{\alpha} - \uuline{\alpha}^\dagger \bigr).
\end{align}
Furthermore, $\mathds{G}_{ij} = \mathds{G}(\mathbf{r}_i, \mathbf{r}_j)$ are the Green functions at positions $\mathbf{r}_{i/j}$ of particle $i$ or $j$. From the expressions of the transmission coefficients it becomes then apparent that an assymetry in the power transferred between the nanoparticles can only exist if there is also an assymetry in the Green's functions, i.e.\ only if $\mathds{G}_{12} \neq \mathds{G}_{21}$. 

\begin{figure}
	\centering
	\includegraphics[width=0.45\textwidth]{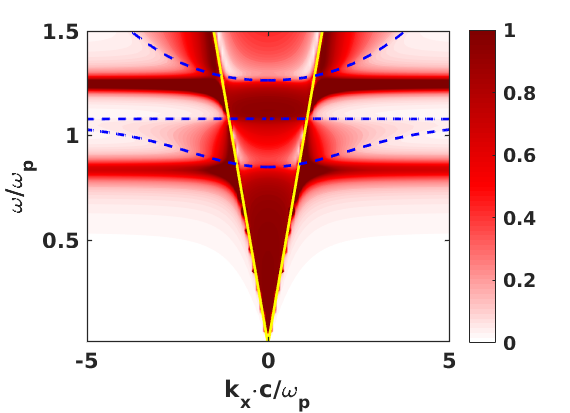}
	\includegraphics[width=0.45\textwidth]{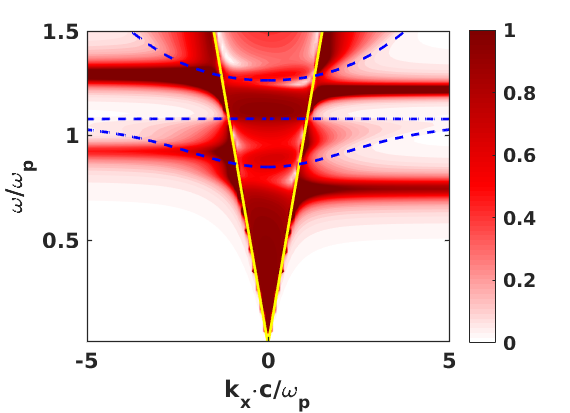}
	\caption{Reflection coefficient $|r_{pp}|$ for $|k_x| \leq k_0$  and $\Im(r_{pp})$ for $|k_x| > k_0$ ($k_y = 0$) without magnetic field (top) and with a magnetic field of 1T (bottom). The solid yellow lines are the ligh lines ($\omega = c k_x$) in vacuum. To better see the surface modes close to the light line in vacuum we have restricted the values to the range $[0:1]$ even though $\Im(r_{pp})$ can have values much larger than $1$. As a guide for the eye we have added dashed blue lines representing the light lines $\omega = c k_x/\sqrt{\Re(\epsilon_3)}$. Without magnetic field these lines correspond to the light line within InSb. In particular, the surface modes exist in this case (top) only in the region with $\Re(\epsilon_3) < 0$.} 
	\label{Fig:ReflectionP}
\end{figure}

Such an assymetry is guaranteed if the material response of the substrate is non-reciprocal, i.e. if at least one of the components of the reflection matrix 
\begin{equation}
  \mathds{R} = \begin{pmatrix} r_{ss} & r_{sp} \\ r_{ps} & r_{pp} \end{pmatrix}
\end{equation}
fulfills the condition $r(\mathbf{k}) \neq r(-\mathbf{k})$. By choosing InSb also for the substrate material and applying a constant magnetic field along the $y$ axis, then for example $r_{pp} (k_x) \neq r_{pp} (-k_x)$~\cite{BrionEtAl1972}. This non-reciprocal behavior can be seen for an InSb substrate in Fig.~\ref{Fig:ReflectionP} where we have plotted $|r_{pp}|$ for the propagating waves with $|k_x| \leq k_0$ and $\Im(r_{pp})$ for the evanescent waves with  $|k_x| > k_0$. We have determined the reflection coefficients by solving the Booker equation~\cite{Chen1981}. The permittivity tensor has the form
\begin{equation}
 \underline{\underline{\rm \epsilon}}=\begin{pmatrix} \epsilon_1 & 0 & -{\rm i}\epsilon_2 \\ 0 & \epsilon_3 & 0 \\ {\rm i}\epsilon_2 & 0 & \epsilon_1 \end{pmatrix},
\label{epsilon}
\end{equation}
with
\begin{equation}
\begin{split}
 \epsilon_1 &= \epsilon_\infty\biggl(1+\frac{\omega_{\rm L}^2-\omega_{\rm T}^2}{\omega_{\rm T}^2-\omega^2-{\rm i}\Gamma\omega}  +\frac{\omega_{\rm p}^2(\omega+{\rm i}\gamma)}{\omega[\omega_{\rm c}^2-(\omega+{\rm i}\gamma)^2]} \biggr), \\
 \epsilon_2 &= \frac{\epsilon_\infty\omega_{\rm p}^2\omega_{\rm c}}{\omega[(\omega+{\rm i}\gamma)^2-\omega_{\rm c}^2]},\\ 
 \epsilon_3 &= \epsilon_\infty\left(1+\frac{\omega_{\rm L}^2-\omega_{\rm T}^2}{\omega_{\rm T}^2-\omega^2-{\rm i}\Gamma\omega}-\frac{\omega_{\rm p}^2}{\omega(\omega+{\rm i}\gamma)} \right).
\end{split}
\end{equation}
The material properties encoded in $\epsilon_{1/2/3}$ for InSb are taken from Ref.~\cite{Zhu2016}, i.e.\ we use for the electronic response $n = 1.07\times10^{17}$\,cm$^{-3}$, $m^* = 1.99\times10^{-32}$\,kg, $\omega_{\rm p} = 3.14\times10^{13}$\,rad/s, and for the phononic response $\gamma = 3.39\times10^{12}$\,rad/s,  $\epsilon_\infty = 15.7$, $\omega_{\rm L} = 3.62\times 10^{13}\,{\rm rad/s}$, $\omega_{\rm T} = 3.39\times10^{13}\,{\rm rad/s}$, and $\Gamma = 5.65\times10^{11}$\,rad/s. Note that if no magnetic field is applied we have $\epsilon_2 = 0$ and  $\epsilon_1 = \epsilon_3$, thus the material becomes isotropic and the permittivity tensor a scalar. In Fig.~\ref{Fig:ReflectionP} it can be seen that in particular the evanescent surface modes traveling along $\pm k_x$ direction are affected by the presence of the magnetic field: The surface-mode resonances are red and blue shifted depending on the sign of $k_x$, as a result of a splitting into two resonances~\cite{BrionEtAl1972}. It is important to note that at the frequencies where the light lines in vacuum and the dashed lines cross there are no surface modes for positive $k_x$ but there is indeed one for negative $k_x$.

To evaluate the heat flux in forward and backward directions we first need the polarizability of the nanoparticles having a radius $R$ which is in dipole approximation (quasi-static limit) given by~\cite{LakhtakiaEtAl1991}
\begin{equation}
  \underline{\underline{\alpha}} = 4\pi R^3(\underline{\underline{\epsilon}}-\mathds{1})(\underline{\underline{\epsilon}}+2\mathds{1})^{-1}.
  \label{alphaerg}
\end{equation}
The dipole approximation is valid if the considered distances and wavelengths are much larger then the radius of the particle. In order to meet this criterion we consider for the numerical calculations particles with 
a radius $R = 5\,{\rm nm}$. In this scenario also the radiation correction is negligible and the single-scattering approximation is valid. Now, we assume that the nanoparticles are also made of InSb to have a good material matching between the nanoparticles and the substrate, but of course the effect demonstrated here also exists for nanoparticles made of another material. Furthermore, we need the Green function $\mathds{G} = \mathds{G}^{(0)} + \mathds{G}^{\text{(sc)}}$ which is the sum of the vacuum Green function $\mathds{G}^{(0)}$ and the scattering part $\mathds{G}^{\text{(sc)}}$ which contains the information of the substrate. Let us now assume that the  particles are placed at positions $\mathbf{r}_1 = (0,0,z)^t$ and $\mathbf{r}_2 = (d,0,z)^t$. Then the vacuum Green function is
given by
\begin{equation}\label{G0}
 \mathds{G}^{(0)} = \frac{e^{ik_0d}}{4\pi k_0^2d^3}\begin{pmatrix}a & 0 & 0\\0 & b & 0\\0 & 0 & b\end{pmatrix},\\
\end{equation}
where $a = 2 - 2 i k_0 d$, $b = k_0^2 d^2 + i k_0 d - 1$ . Of course the vacuum 
Green function is reciprocal and depends only on the interparticle distance, i.e.\ $\mathds{G}^{(0)}_{12} = \mathds{G}^{(0)}_{21}$. 
Now, the scattering part of the Green function can be expressed as
\begin{equation}
  \mathds{G}_{12/21}^{\text{(sc)}} = \int_{-\infty}^{+\infty} \!\! \frac{\rd k_x}{2 \pi} \int_{-\infty}^{+\infty} \!\! \frac{\rd k_y}{2 \pi} \, \re^{\mp \ri k_x d} \mathds{G}^{\text{(sc)}}(k_x,k_y; z)
\end{equation} 
with
\begin{equation}
  \mathds{G}^{\text{(sc)}}(k_x,k_y; z) = \frac{2 \ri \re^{\ri \gamma_0 z}}{2 \gamma_0} \sum_{i,j = p,s} r_{ij} \mathbf{a}_i^+ \otimes \mathbf{a}_j^-.
\end{equation}
The polarization vectors are defined as $\mathbf{a}_s^\pm = \frac{1}{\kappa}(k_y, -k_x,0)^t$ and $\mathbf{a}_p^\pm = \frac{1}{\kappa k_0} (\mp k_x \gamma_0,\mp k_y \gamma_0, \kappa^2)$ where $\kappa = \sqrt{k_x^2 + k_y^2}$ and $\gamma_0 = \sqrt{k_0^2 - \kappa^2}$. From these expressions it is obvious that $ \mathds{G}_{12}^{\text{(sc)}} =  \mathds{G}_{21}^{\text{(sc)}}$ only if the integrand fulfills $\mathds{G}^{^{\text{(sc)}}}(k_x,k_y; z) = \mathds{G}^{^{\text{(sc)}}}(-k_x,k_y; z)$. As we have seen before, when a magnetic field is applied we have $r_{pp} (k_x) \neq r_{pp} (-k_x)$ and hence $\mathds{G}^{\text{(sc)}}(k_x,k_y; z) \neq \mathds{G}^{sc}(-k_x,k_y; z)$ and $\mathds{G}_{12}^{\text{(sc)}} \neq \mathds{G}_{21}^{\text{(sc)}}$.

\begin{figure}
	\centering
	\includegraphics[width=0.45\textwidth]{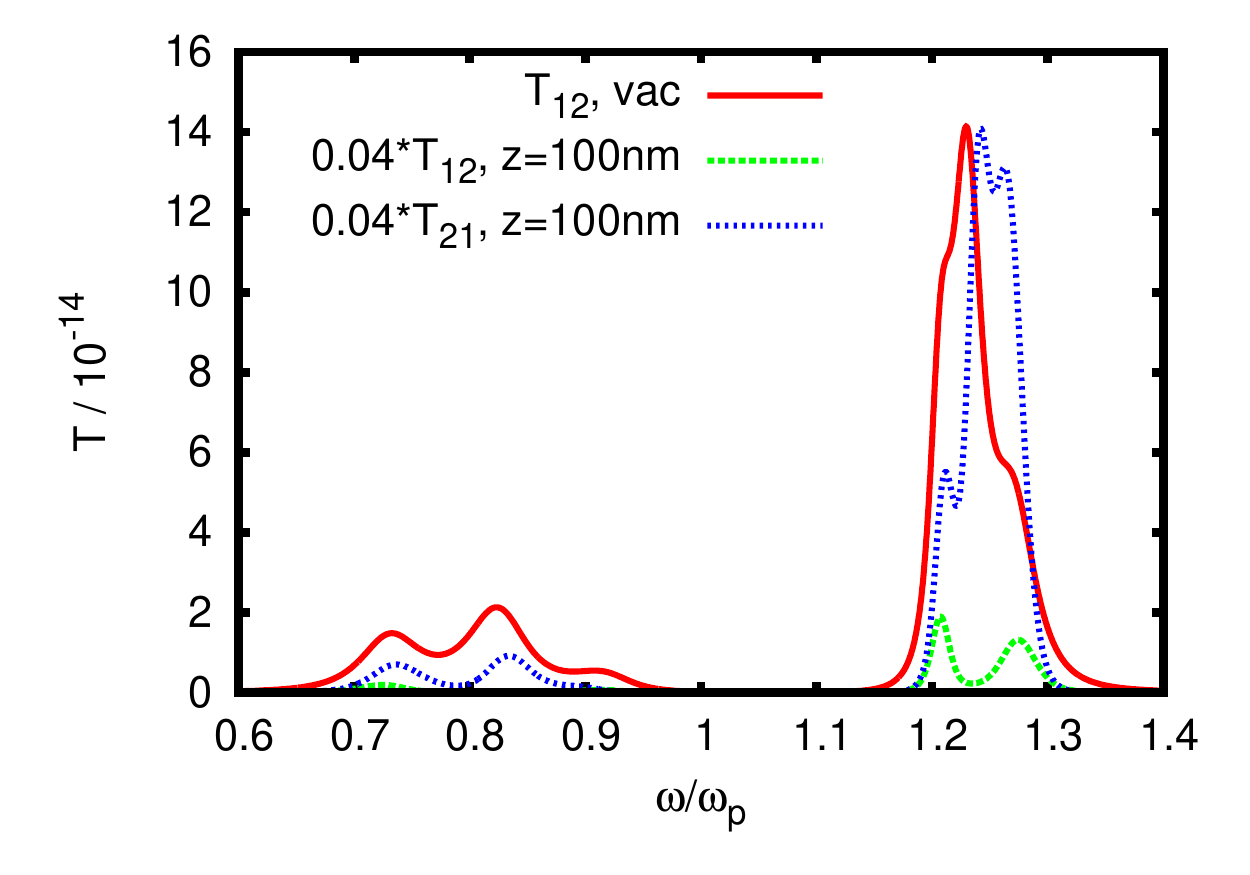}
	\caption{The solid red line gives the transmission coefficients $\mathcal{T}_{12} = \mathcal{T}_{21}$ for two InSb nanoparticles in vacuum with $d = 1\,\mu$m and $B = 1\,$T. The dashed green and blue lines correspond to the scaled transmission coefficents $\mathcal{T}_{12}$ and $\mathcal{T}_{21}$, respectively, in the configuration where the two nanoparticles are at distance $z = 100\,{\rm nm}$ from a InSb substrate.}
	\label{Fig:T12}
\end{figure}

\begin{figure}
	\centering
	\includegraphics[width=0.45\textwidth]{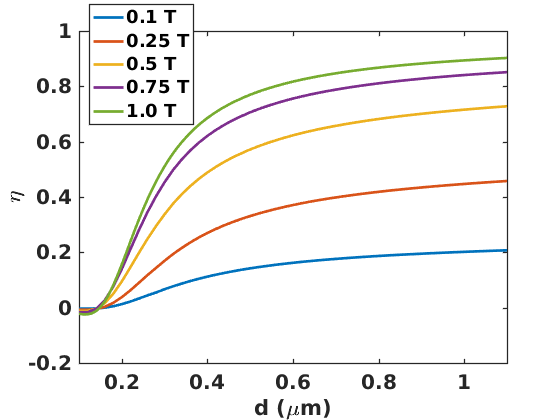}
 	\includegraphics[width=0.45\textwidth]{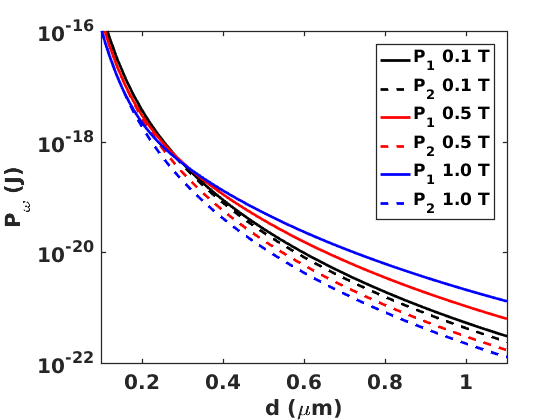}
	\caption{Top: Rectification coefficient $\eta$ for the heat flux between two nanoparticles above a InSb substrate with distance $z = 100\,{\rm nm}$ as function of the interparticle distance $d$ and the magnitude of the magnetic field pointing along the $y$ axis. Bottom: Corresponding absolute power $P_1$ and $P_2$ for a choice of magnetic fields.}
	\label{Fig:Rect}
\end{figure}

In the following we will use the above definitions to evaluate the power flow in backward and forward directions. In the numerical calculations we neglect the depolarization terms $r_{sp}$ and $r_{ps}$ because they turn out to be negligibly small. To be more specific, the main effect results from the term proportional to $r_{pp}$. In Fig.~\ref{Fig:T12} it can be seen that when considering two InSb nanoparticles with an interparticle distance of $1\,\mu$m in vacuum and when applying a magnetic field we have three resonances in $\mathcal{T}_{12} = \mathcal{T}_{12}$, associated with the dipolar resonances with magnetic quantum numbers $m = 0, \pm 1$. If the particles are now at distance $z = 100\,{\rm nm}$ above a InSb substrate, we clearly see that the condition $\mathcal{T}_{12} \neq \mathcal{T}_{21}$ is realized. Moreover, we observe the transmission coefficents are larger by a factor of about 25, which means that due to the coupling to the surface modes of the sample the transmitted power transfer increases, as already whitnessed in the isotropic case~\cite{DongEtAl2018,MessinaEtAl2018}. Second, it can be seen that the middle resonance in $\mathcal{T}_{12}$ is suppressed. This is due to the fact that in this spectral region no surface modes propagating in the positive $x$ direction exist. This is coherent with the plots of $r_{pp}$ shown in Fig.~\ref{Fig:ReflectionP}. Furthermore, the resonances in $\mathcal{T}_{21}$ are stronger than in $\mathcal{T}_{12}$, as a result of the longer propagation length of the surface modes which travel in the negative $x$ direction with respect to the surface modes traveling in the positive $x$ direction. As a consequence, the heat flux from particle $1$ to particle $2$ is smaller than from $2$ to $1$, i.e. we observe a rectification of the heat flux. In order to quantify this rectification we introduce the rectification coefficient 
\begin{equation}
  \eta = \frac{P_1 - P_2}{P_1}, 
\end{equation}
where $P_1$ is the power received by particle $1$ in the backward case where $T_1 = T_b = 300\,{\rm K}$ and $T_2 = 350\,{\rm K}$ and $P_2$ is the power received by particle $2$ in the forward case where $T_2 = T_b = 300\,{\rm K}$ and $T_1 = 350\,{\rm K}$. Since we know already from the discussion of the transmission coefficients that at least for large interparticle distances $P_1 > P_2$, our rectification coefficent is by definition smaller than $1$. In Fig.~\ref{Fig:Rect} we show $\eta$ as function of the interparticle distance $d$ and the magnitude of the magnetic field $B$. It is apparent that for distances $d$ smaller or close to particle-substrate distance $z = 100\,{\rm nm}$ there is practically no rectification because in this case the interaction between the particles is stronger than the interaction between the particles and the surface. On the contrary, for distances $d \gg z = 100\,{\rm nm}$ the interaction between the particles is mainly due to the coupling to the surface modes of the substrate leading to the rectification which can be easily larger than 50\%.

In conclusion, we have introduced here a new physical mechanism to rectify the radiative heat flux exchanged in near-field regime between two hot bodies when there are located close to non-reciprocal media. This mechanism is based on the asymmetry with respect their propagation direction of evanescent waves confined at the surface of these media. In particular, when these media support surface waves in the Planck window where the two bodies interact we have demonstrated that the rectification coefficient can reach very large values opening the door to true thermal diode. Thanks to the diversity of non-reciprocal media these familly of diodes could work over a broad temperature ranges. By using magneto-optical materials we have also shown that the rectification coefficient can be actively tuned by changing the magnitude of a magnetic field applied on the systems. We think that this technology could find broad applications in the field of thermal management at nanoscale.
%

\acknowledgments
\noindent

S.-A.\ B. acknowledges support from Heisenberg Programme of the Deutsche Forschungsgemeinschaft (DFG, German Research Foundation) under the project No. 404073166.

\end{document}